\documentclass[usenatbib,nofootinbib,twocolumn,superscriptaddress]{revtex4}

\usepackage{graphicx}
\usepackage{amssymb}
\usepackage{epstopdf}
\usepackage{aas_macros}
\usepackage{color}

\newcommand{\cpm}{{\small CUBEP$^3$M}}
\newcommand{\hfit}{{\small HALOFIT}}
\newcommand{\camb}{{\small CAMB}}

\newcommand{\mpch}{\mbox{Mpc}/\mbox{h}}
\newcommand{\hmpc}{\mbox{h}/\mbox{Mpc}}
\newcommand{\ev}{{\rm eV}}
\newcommand{\kms}{{\rm km}/{\rm s}}
\newcommand{\tdel}{T_\delta}
\newcommand{\tvel}{T_v}

\begin{document}

\title{Precision reconstruction of the dark matter-neutrino relative
  velocity from N-body simulations.}

\author{Derek Inman} \affiliation{Canadian Institute for Theoretical
  Astrophysics, University of Toronto, 60 St. George St., Toronto, ON
  M5S 3H8, Canada} \affiliation{Department of Physics, University of
  Toronto, 60 St. George, Toronto, ON M5S 1A7, Canada}

\author{J.D. Emberson} \affiliation{Canadian Institute for Theoretical
  Astrophysics, University of Toronto, 60 St. George St., Toronto, ON
  M5S 3H8, Canada} \affiliation{Department of Astronomy and
  Astrophysics, University of Toronto, 50 St. George, Toronto, ON M5S
  3H4, Canada}

\author{Ue-Li Pen} \affiliation{Canadian Institute for Theoretical
  Astrophysics, University of Toronto, 60 St. George St., Toronto, ON
  M5S 3H8, Canada} \affiliation{Canadian Institute for Advanced
  Research, CIFAR Program in Gravitation and Cosmology, Toronto,
  Ontario M5G 1Z8, Canada}

\author{Alban Farchi} \affiliation{\'{E}cole polytechnique, Route de
  Saclay, 91120 Palaiseau, France}

\author{Hao-Ran Yu} \affiliation{Canadian Institute for Theoretical
  Astrophysics, University of Toronto, 60 St. George St., Toronto, ON
  M5S 3H8, Canada} \affiliation{Kavli Institute for Astronomy and
  Astrophysics, Peking University, Beijing 100871, China}

\author{Joachim Harnois-D\'{e}raps} \affiliation{Department of Physics
  and Astronomy, The University of British Columbia, 6224 Agricultural
  Road, Vancouver, B.C., V6T 1Z1, Canada} 

\begin{abstract}
  Discovering the mass of neutrinos is a principle goal in high energy
  physics and cosmology.  In addition to cosmological measurements
  based on two-point statistics, the neutrino mass can also be
  estimated by observations of neutrino wakes resulting from the
  relative motion between dark matter and neutrinos.  Such a detection
  relies on an accurate reconstruction of the dark matter-neutrino
  relative velocity which is affected by non-linear structure growth
  and galaxy bias.  We investigate our ability to reconstruct this
  relative velocity using large N-body simulations where we evolve
  neutrinos as distinct particles alongside the dark matter.  We find
  that the dark matter velocity power spectrum is overpredicted by
  linear theory whereas the neutrino velocity power spectrum is
  underpredicted.  The magnitude of the relative velocity observed in
  the simulations is found to be lower than what is predicted in
  linear theory.  Since neither the dark matter nor the neutrino
  velocity fields are directly observable from galaxy or 21 cm
  surveys, we test the accuracy of a reconstruction algorithm based on
  halo density fields and linear theory.  Assuming prior knowledge of
  the halo bias, we find that the reconstructed relative velocities
  are highly correlated with the simulated ones with correlation
  coefficients of 0.94, 0.93, 0.91 and 0.88 for neutrinos of mass
  0.05, 0.1, 0.2 and 0.4 $\ev{}$.  We confirm that the relative
  velocity field reconstructed from large scale structure observations
  such as galaxy or 21 cm surveys can be accurate in direction and,
  with appropriate scaling, magnitude.
\end{abstract}

\maketitle

\begin{section}{Introduction}
  \label{sec:introduction}

  Despite extensive research in the particle physics and cosmology
  communities, many properties of neutrinos remain elusive.  For
  instance, neutrino oscillation experiments \citep{bib:Fogli2012}
  have accurately measured the mass-squared splittings between
  neutrino species, but individual neutrino masses have yet to be
  measured.  It is also unknown whether the neutrino masses follow a
  {\em normal} hierarchy in which there are two light neutrinos and a
  single heavy one or an {\em inverted} hierarchy with the opposite
  configuration.  Moreover, it is still unknown whether neutrinos are
  Dirac or Majorana fermions.
   
  Cosmological techniques for determining neutrino masses are
  currently insensitive to individual neutrinos and instead constrain
  the sum of all neutrino masses. For instance, cosmic microwave
  background (CMB) observations made by the Plank satellite place
  $\sum m_\nu < 0.194\ \ev$ \citep{bib:Planck2015}. Recently, a new
  technique for constraining neutrino mass using large-scale velocity
  fields was proposed in \citep{bib:Zhu2013,bib:Zhu2014}.  Neutrinos
  and dark matter are expected to have a relative velocity arising due
  to the free streaming of neutrinos over large scales.  As neutrinos
  bulk flow over large scale structures they become focussed into
  wakes.  Such downstream overdensities introduce a unique dipole
  distortion in the matter field in the direction of the neutrino flow
  which could be observed via either direct lensing of the wake or
  through a dipole component of the correlation function.

  Unlike other probes of cosmological neutrinos, this method is
  expected to be background free and only relies on knowledge of the
  relative velocity field.  Determining velocity fields directly is
  particularly challenging even for luminous matter and certainly
  impossible for neutrinos.  However, the relative velocity is
  predicted to be coherent over several megaparsecs.  We therefore
  expect linear theory to be accurate enough to allow for a
  reconstruction of the velocity field from the easier to obtain
  matter density field.

  Our goal is to quantify the accuracy of this linear reconstruction
  when non-linear structure formation, which affects both the density
  and velocity fields, is taken into account.  We furthermore wish to
  understand whether the reconstruction procedure is robust when only
  a {\it tracer} of the dark matter field is used.  To achieve this we
  use large cosmological simulations.  Neutrinos have been implemented
  in a variety of ways within the framework of N-body simulations: (i)
  \citep{bib:Brandbyge2009} used a grid-based approach where an
  additional neutrino density field is evolved alongside N-body dark
  matter; (ii) \citep{bib:Brandbyge:2010} employed a hybrid method
  where neutrinos start as a grid and are converted to particles as
  their energy decreases; (iii) \citep{bib:Bird2012} evolved neutrinos
  as distinct N-body particles; (iv) \citep{bib:AliHaimoud2013}
  computed the neutrino linear response alongside the evolving dark
  matter. In general, the grid-based approaches have been unable to
  resolve non-linear neutrino structure formation while particle-based
  approaches are hindered by the requirement that many neutrino
  particles are needed to reduce Poisson noise on small scales. In
  this work, we adopt the particle based approach since an accurate
  computation of non-linear neutrino dynamics is a main focus of our
  work.

  In \S \ref{sec:theory} we discuss our implementation of neutrino
  particles into the cosmology code \cpm{}
  \citep{bib:HarnoisDeraps2013} and our method for computing density
  and velocity fields.  In \S \ref{sec:results} we present the results
  of our simulations and analyse the accuracy of various
  reconstruction methods.  In \S \ref{sec:discussion} we discuss a
  practical procedure to estimate cosmic velocity fields from density
  tracers.

\end{section}

\begin{section}{Theory and Implementation}
  \label{sec:theory}

  \begin{subsection}{Neutrino N-body Particles in \cpm{}}
    \label{ssec:neutrinos}
  
    Initial neutrino positions are generated separately from dark
    matter using the same Gaussian noise map. We use neutrino density
    transfer functions, $\tdel$, computed via \camb{}
    \citep{bib:Lewis}.  The initial neutrino velocity is composed of
    two parts: a linear component (analogous to the Zel'dovich
    velocity) plus a random thermal component.  For the linear
    component, we first compute the linear neutrino velocity transfer
    function, $\tvel$, via the continuity equation under the
    assumption that initial conditions are adiabatic and velocities
    are linear (e.g. $\delta(k,z) = \tdel(k,z) \delta_i(k)$ and
    $\vec{v}(\vec{k},z) = \tvel(k,z) \delta_i(k) \hat{k}$ for an
    initial perturbation $\delta_i(k)$):
    \begin{equation}
      \dot{\delta} + \frac{1}{a}\vec{\nabla}\cdot\vec{v} = 0 \rightarrow \tvel = - i \frac{H}{k} \frac{\tdel(z+\delta z) - \tdel(z-\delta z)}{2 \delta z},
      \label{eq:veltransfer}
    \end{equation}
    where we convert time derivatives to redshift derivatives and
    evaluate numerically using a spacing $\delta z = 0.1$. We have
    checked that the transfer functions computed via
    Eq. \ref{eq:veltransfer} are in good agreement with those produced
    by the {\small CLASS} code \citep{bib:Blas} in Newtonian
    gauge\footnote{The {\small CAMB} density transfer functions are in
      the synchronous gauge whereas the velocity transfer function we
      desire are in the longitudinal Newtonian gauge.  However, the
      gauge transformation terms are proportional to the time
      derivatives of the Newtonian potentials which we already ignore
      in the continuity equation.}.

    From this velocity transfer function, we compute a velocity
    potential, $\phi_v(k)$, such that
    $\vec{v}(k) = i \vec{k} \phi_v(k) = (\tvel/\tdel)\delta \hat{k}$.
    When combined with Eq. \ref{eq:veltransfer} this yields:
    \begin{equation}
      \phi_v(k) = - \frac{H}{k} \frac{\tdel(z+\delta z) - \tdel(z-\delta z)}{2 \delta z} \frac{\delta}{\tdel}.
      \label{eq:velpotential}
    \end{equation}
    This potential is then Fourier transformed and a two-sided finite
    difference is taken to obtain the linear velocity.  Using a
    real-space gradient reduces the number of Fourier transforms to be
    computed and is consistent with our calculation of the
    displacement field.

    The random component of the velocity is computed via the
    cumulative distribution function, ${\rm CDF}[v,\beta]$, which
    follows from the relativistic Fermi-Dirac distribution,
    ${\rm PDF}[v,\beta]$, for neutrinos:
    \begin{eqnarray}
      {\rm PDF}[v,\beta] &= & \frac{v^2}{e^{m_\nu v/kT}+1} = \frac{v^2}{e^{v\beta}+1} \nonumber \\
      {\rm CDF}[v, \beta] &=& \frac{1}{{\rm CDF}[\infty,1]} \beta^3 \int_o^v
                              {\rm PDF}[u,\beta] dv \nonumber   \\
                         &=&\frac{1}{{\rm CDF}[\infty,1]} \int_o^{u=v\beta} {\rm PDF}[u,1] du \nonumber \\
                         &=&{\rm CDF}[u,1]
                             \label{eq:fermicdf}
    \end{eqnarray}
    where $m_\nu$ and $T$ are neutrino mass and temperature,
    respectively, and $\beta \equiv m_\nu/kT$.  Our numerical
    evaluation of the $\rm CDF$ gives a maximum particle speed of
    $0.013 \left(0.2\:\ev/m\right) (1+z^i)c$.  Neutrinos in the mass
    regime we are interested in are relativistic at the redshift for
    which dark matter initial conditions are generated ($z_c = 100$):
    \begin{equation}
      \langle v \rangle = \frac{\int_0^\infty v{\rm PDF}[v,\beta]
        dv}{\int_0^\infty {\rm PDF}[v,\beta] dv} \approx 800
      \left(\frac{0.2\ \ev}{m_\nu}\right) (1+z)\ \kms.
    \end{equation}
    This thermal motion would dominate the time step constraining the
    maximum distance a particle may travel, making the simulation
    impractically slow.  To circumvent this issue we evolve the dark
    matter in isolation to a lower redshift, $z_\nu \sim 10$, at which
    point neutrinos are added and the two components evolve together.
  
    During their subsequent evolution, dark matter and neutrino
    particles are treated identically except for their masses, which
    are weighted by their energy fractions as well as number ratio:
    \begin{equation}
      m_i = \frac{\Omega_i}{\Omega_m}\frac{N_g}{N_i},
      \label{eq:pmass}
    \end{equation}
    where $\Omega_i$ is the energy fraction of species $i$, $\Omega_m$
    is the total matter energy fraction, $N_g$ is the number of cells
    in the simulation grid, and $N_i$ is the number of particles of
    species $i$.  These masses are used when adding particles to the
    grid for the computation of the long-range gravitational force as
    well as the short-range pairwise force.  The particle type is
    distinguished within the code using 1 byte particle identification
    tags.

  \end{subsection}

  \begin{subsection}{Density and Velocity Fields}
    \label{ssec:velocity}
  
    We compute dark matter, neutrino, and halo density fields using a
    standard cloud-in-cell interpolation method for both dark matter
    and neutrinos. Computing velocity fields from particle-based
    simulations has only recently been studied in depth. This may be
    related to the ambiguity associated with defining a velocity field
    from a sample of point particles. Unlike quantities such as mass
    or momentum, the velocity of a particle cannot be simply added to
    a grid. The most obvious method for generating a velocity field is
    to divide a gridded momentum field by its corresponding density
    field. However, within void regions it is possible that empty
    cells exist for which no well-defined velocity can be
    assigned. Alternatively, one may define the velocity at a given
    grid cell to be the average velocity of the $N_{\rm near}$ nearest
    particles about this point.  The application of the nearest
    particle method was studied by \cite{bib:Zhang2014} and
    \cite{bib:Zheng2014} where it was found that the velocity power is
    suppressed for low particle number densities,
    $ n<1\ (\mpch)^{-3}$, due to the sampling procedure.  In our
    simulations we use high number densities,
    $n_{\rm dm} \sim 10\ (\mpch)^{-3}$, and therefore do not expect
    this effect to be significant.  More advanced methods for
    computing velocity fields exist such as phase-space interpolation
    discussed in \citep{bib:Pueblas} and more recently in
    \citep{bib:Hahn2014}.
  
    In what follows we compute the velocity fields of dark matter and
    neutrinos in different ways.  For dark matter, we adopt the
    nearest particle method and take the $N_{\rm near} = 1$ nearest
    particle about the centre of each cell using the same grid
    resolution as neutrinos.  We have found that the nearest particle
    method can also be used for neutrinos albeit with a much larger
    $N_{\rm near} = 64$ to smooth the field on small scales. However,
    searching over this many particles is a computationally expensive
    task.  For neutrinos we therefore employ the approach of dividing
    their momentum field by their density field on grids coarsened so
    that there is always at least one neutrino per cell. This is
    possible since neutrinos are rather homogeneously distributed and
    form voids to a lesser extent than dark matter.

    We treat the velocity fields obtained from the nearest particle
    and momentum methods as faithful tracers of the actual field.
    However, these fields are not comparable to observational data
    since neither dark matter nor neutrino velocities can be directly
    measured.  For this purpose we reconstruct velocity fields from
    density fields using linear theory:
    \begin{equation}
      \vec{v} = \frac{\tvel}{\tdel}\frac{\vec{k}}{k}\delta,
      \label{eq:vrec}
    \end{equation}
    where we use dark matter and halo density fields separately for
    $\delta$ (although with the same $\tdel$).  In what follows we
    treat halos as point particles of unit mass in order to represent
    the information available through galaxy surveys.

    Poisson noise is a severe hindrance for the neutrinos (and to a
    far lesser extent dark matter) due to their large thermal
    velocities.  For density fields it is possible to subtract out the
    Poisson noise but this is not possible for velocities.  To remove
    this noise in the density and velocity auto-power spectra we
    instead randomly divide each species of particles into two groups
    for which separate fields -- $f_i^1$ and $f_i^2$ -- are
    computed. The dimensionless power spectrum is then computed as
    \begin{equation}
      \Delta^2_{ii}(k) = \frac{k^3}{2\pi^2} \langle f_i^{1}f_i^2 \rangle,
      \label{eq:deltaij}
    \end{equation}
    where index $i = c, \nu, h$ denotes dark matter, neutrinos, and
    halos, respectively. This method effectively removes noise as the
    noise in each group is uncorrelated and cancels out in the cross
    term.  We note that this method is only used when computing
    auto-power since a cross-power $\Delta^2_{ij}$ with $i \neq j$
    automatically washes out noise that is uncorrelated between
    separate species.

    The accuracy of the reconstructed field is measured using a
    correlation coefficient:
    \begin{equation}
      r_{ij}(k) = \frac{\Delta^2_{ij}(k)} { \sqrt{\Delta^2_{ii}(k)\Delta^2_{jj}(k)} }
      \label{eq:rij} 
    \end{equation}
    where $\Delta^2_{ij}$ is the cross power spectrum between species
    $i=c,h,\nu \textrm{ or rel}$ using reconstruction method
    $\mbox{sim}, \mbox{Rec DM}, \mbox{Rec HA}$ (nearest
    particle/momentum, Eq. \ref{eq:vrec} with CDM and
    Eq. \ref{eq:vrec} with haloes respectively) and species $j$ (with
    potentially a different reconstruction method).  We also define
    the integrated correlation coefficient as:
    \begin{equation}
      r_{ij} = \frac{\int \Delta^2_{ij} \frac{dk}{k}}{\sqrt{\int \Delta^2_{ii} \frac{dk}{k}} \sqrt{\int \Delta^2_{jj} \frac{dk}{k}}}
      \label{eq:intrij}
    \end{equation}
    which no longer depends on wavenumber.

  \end{subsection}

\end{section}

\begin{section}{Results}
  \label{sec:results}

  In this Section we present the results for a suite of four
  simulations of dark matter and neutrinos.  We simulate neutrinos of
  mass $m_\nu = 0.4, 0.2, 0.1$ and $0.05\:\ev{}$.  Each simulation
  contains $N_c = 1536^3$ dark matter particles and $N_\nu = 3072^3$
  neutrino particles within a periodic box of side length
  $L = 500\ \mpch$. In each case dark matter is started from an
  initial redshift $z_c = 100$ and gravitational forces are softened
  below the scale $r_{\rm soft} = 24$ kpc/$h$.  Neutrinos are added in
  at redshift $10$ for all species except $0.05\:\ev{}$ which we add
  at redshift $5$.  We assume a base cosmology compatible with Planck
  results: $\Omega_b = 0.05$, $\Omega_c = 0.27$, $\sigma_8 = 0.83$,
  $n_s = 0.96$, $h = 0.67$, and compute
  \begin{equation}
    \Omega_{\nu} = \frac{m_\nu}{93.14\ h^2}
    \label{eq:omnu}
  \end{equation}
  as in \citep{bib:Mangano2005}. We hold $\Omega_b$ and $\Omega_c$
  fixed in each simulation and maintain a flat universe by adjusting
  $\Omega_\Lambda = 1 - \Omega_m = 1 - \Omega_b - \Omega_c -
  \Omega_\nu$.
  In what follows we mainly investigate a fiducial simulation with
  $m_\nu = 0.2$ eV.  We label our simulations based on neutrino mass
  with S05, S1, S2, and S4 denoting the simulations with
  $m_\nu = 0.05, 0.1, 0.2$, and $0.4\:\ev{}$ respectively.
  
  Halo catalogues are generated for each simulation at $z = 0$ using a
  spherical overdensity algorithm that considers all halos with at
  least 100 dark matter particles. This corresponds to a minimum halo
  mass of $3 \times 10^{11}\ M_\odot/h$. Recall, however, that we
  assign each halo unit mass when constructing halo density fields in
  order to emulate the information available in galaxy surveys. In
  what follows, density and velocity fields for dark matter,
  neutrinos, and halos are computed on uniform rectilinear grids
  containing $1536^3$ mesh cells.

  \begin{subsection}{Density}
    \label{ssec:density}
    \begin{figure*}[htbp]
      \begin{center}
        \includegraphics[width=0.9\textwidth]{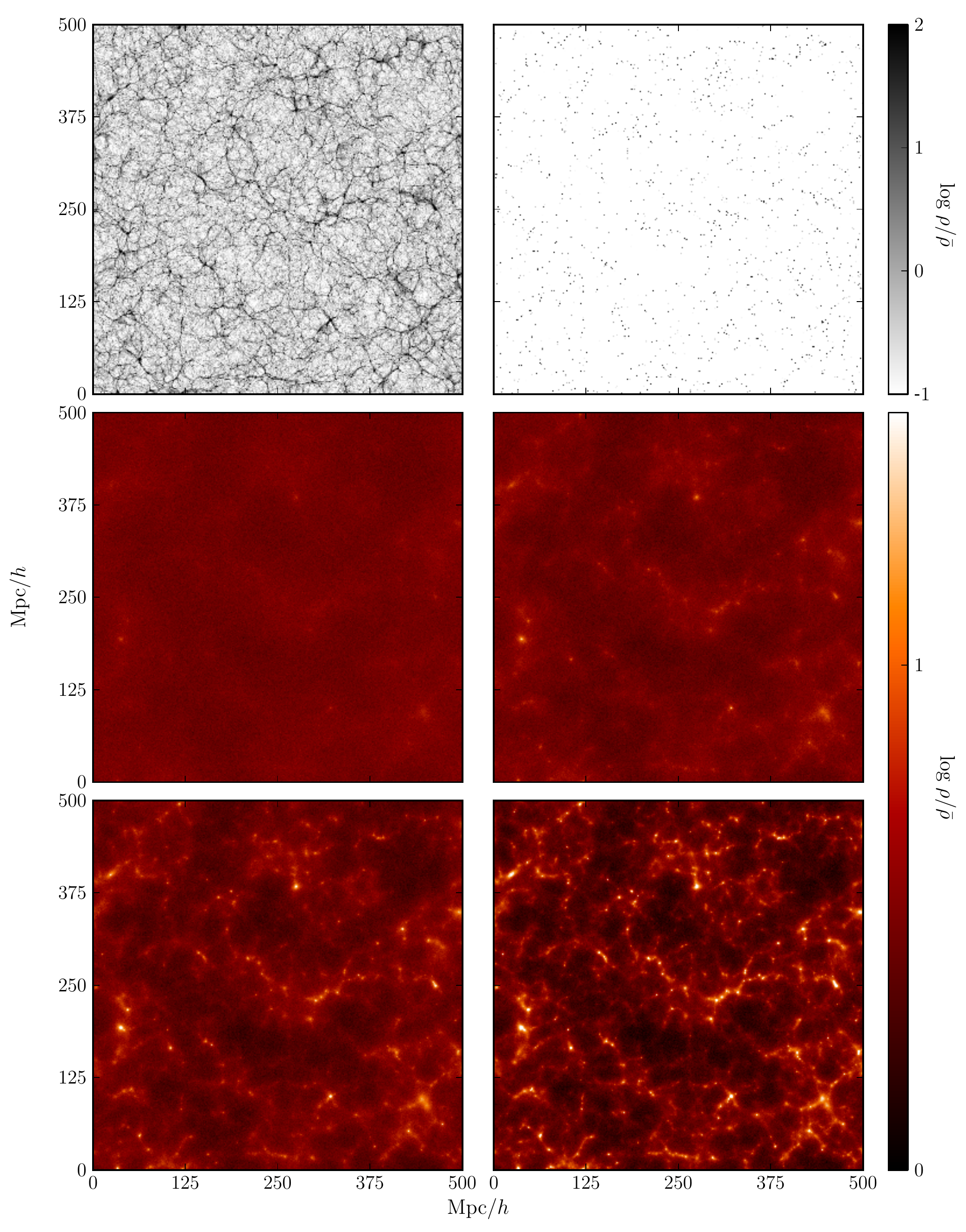}
        \caption{Density slices of equal width $500\ \mpch{}$ and
          thickness $1.3\ \mpch{}$ from various simulations at
          $z = 0$.  The top row shows dark matter (left) and halo
          (right) density slices from the 0.2\ \ev{} neutrino
          simulation. The middle row compares neutrino density slices
          from the 0.05 (left) and 0.1 (right)\ \ev{} simulations
          while the bottom row shows the 0.2 (left) and 0.4 (right)\
          \ev{} simulations.  It is easy to see by eye that the dark
          matter and neutrino density fields are highly correlated and
          that heavier neutrinos cluster more than lighter ones.}
        \label{fig:denslice}
      \end{center}
    \end{figure*}
    
    Fig. \ref{fig:denslice} compares slices of the dark matter and
    halo density fields at $z = 0$ from simulation S2 to the neutrino
    density fields from simulations S05, S1, S2, and S4.  It is easy
    to see that the neutrino density fields are correlated with the
    dark matter density field albeit with much less clumping in the
    former than the latter as evidenced by their respective colour
    bars. In addition, we see that higher mass neutrinos tend to clump
    more than lower mass neutrinos as they are more influenced by the
    underlying dark matter distribution due to their lower thermal
    velocities.

    \begin{figure}[htbp]
      \begin{center}
        \includegraphics[width=0.5\textwidth]{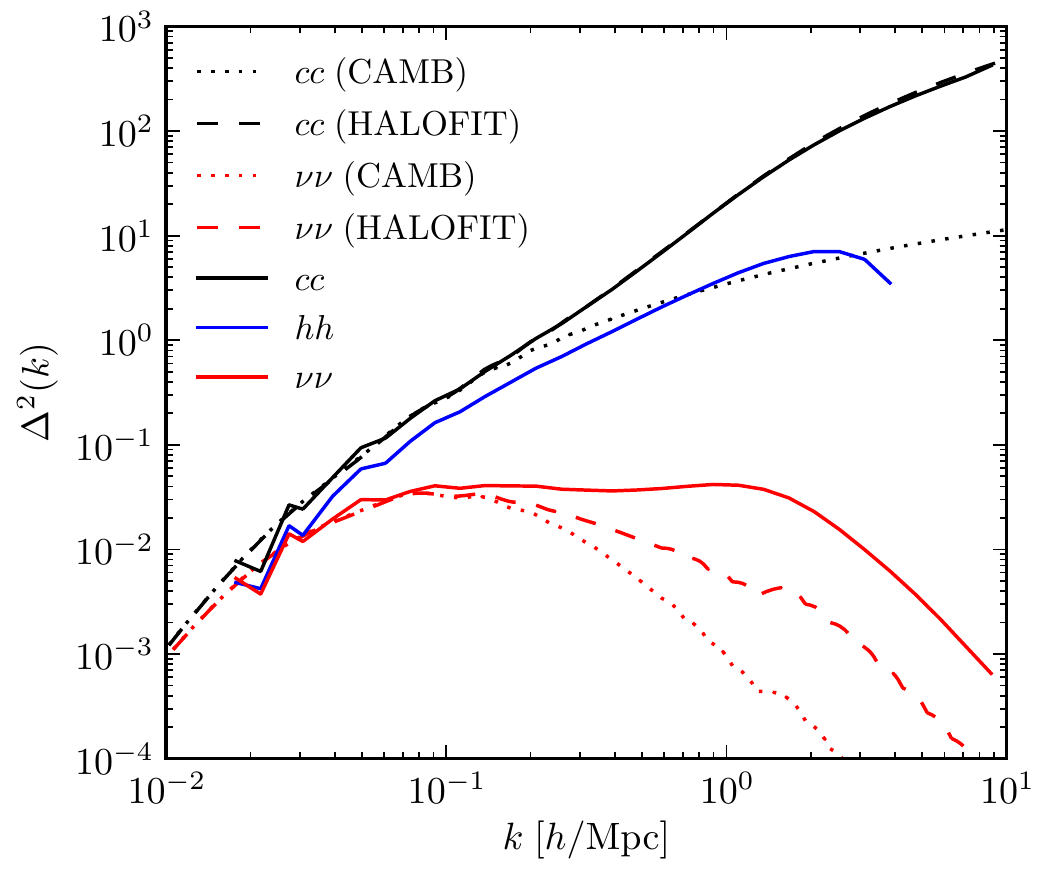}
        \caption{The dimensionless matter power spectra at $z = 0$ for
          dark matter (solid black line), halos (solid blue line) and
          neutrinos (solid red line) from S2. Shot noise has been
          removed by computing the cross-spectrum between two randomly
          chosen groups for each species.  Also plotted are the linear
          and non-linear dark matter (dotted black and dashed black
          lines) and neutrino (dotted red and dashed red lines) power
          spectra. Note that there is a small numerical artifact in
          the linear neutrino transfer function just above
          $k = 1\:\hmpc$ that should be ignored. }
        \label{fig:denpow}
      \end{center}
    \end{figure}

    Fig. \ref{fig:denpow} shows the dimensionless power spectra for
    dark matter, halos, and neutrinos at $z = 0$ from S2.  Also
    plotted are theoretical predictions for dark matter and neutrinos,
    which are computed via
    \begin{equation}
      \Delta^2_i(k) = \frac{k^3}{2\pi^2}  P_m\left(\frac{T_i}{T_m}\right)^2,
      \label{eq:PNL}
    \end{equation}    
    where $T_i$ is the linear transfer function for species $i$, $T_m$
    is the total matter linear transfer function, and $P_m$ is either
    the linear (computed from \camb{}) or the non-linear (computed
    from \hfit{}) total matter power spectrum. We first note that the
    group cross-correlation method we employ effectively removes the
    shot noise allowing us to understand statistical properties even
    of the noisy neutrino density field.  We find that the dark matter
    power spectrum agrees well with the non-linear prediction up to
    large $k$.  The neutrino power spectrum, on the other hand, is
    significantly enhanced on small scales compared to the theoretical
    curve.  This trend was previously observed by
    \citep{bib:AliHaimoud2014} and is yet to be understood.
    
    \begin{figure}[htbp]
      \begin{center}
        \includegraphics[width=0.5\textwidth]{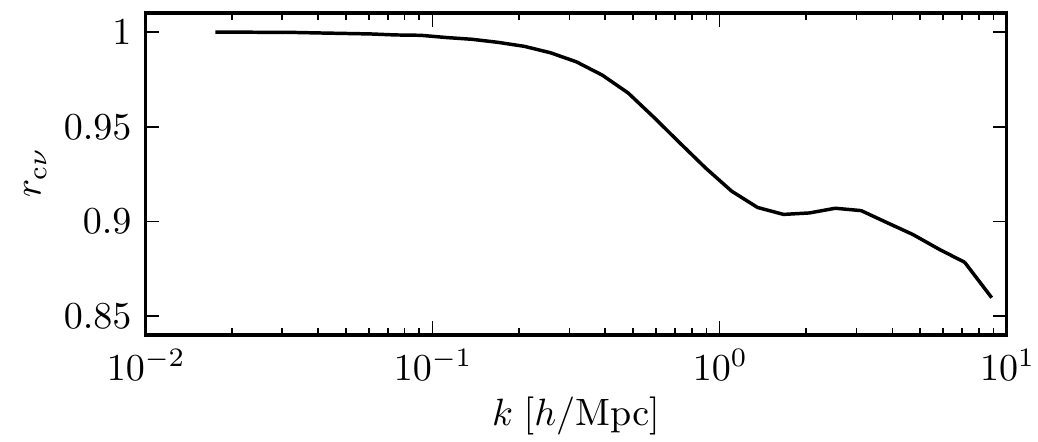}
        \caption{The dark matter-neutrino cross correlation
          coefficient at $z = 0$ from S2. As expected, neutrinos are
          highly correlated with dark matter over a large range of
          scales. }
        \label{fig:dencorr}
      \end{center}
    \end{figure}
    
    Despite their enhanced power on small scales, neutrinos remain
    highly correlated with the dark matter density field, as was
    qualitatively discussed with Fig. \ref{fig:denpow}.  More
    quantitatively, Fig. \ref{fig:dencorr} shows the $z = 0$
    cross-correlation coefficient between dark matter and neutrinos
    from S2 as a function of wavenumber. We find that neutrinos
    exhibit $r_{c\nu} \gtrsim .90$ correlation with dark matter on all
    scales $k < 1\ \hmpc$ and achieve $r_{c\nu} \sim .85$ down to the
    smallest scales resolved in the simulation.
    
    \begin{figure}[htbp]
      \begin{center}
        \includegraphics[width=0.5\textwidth]{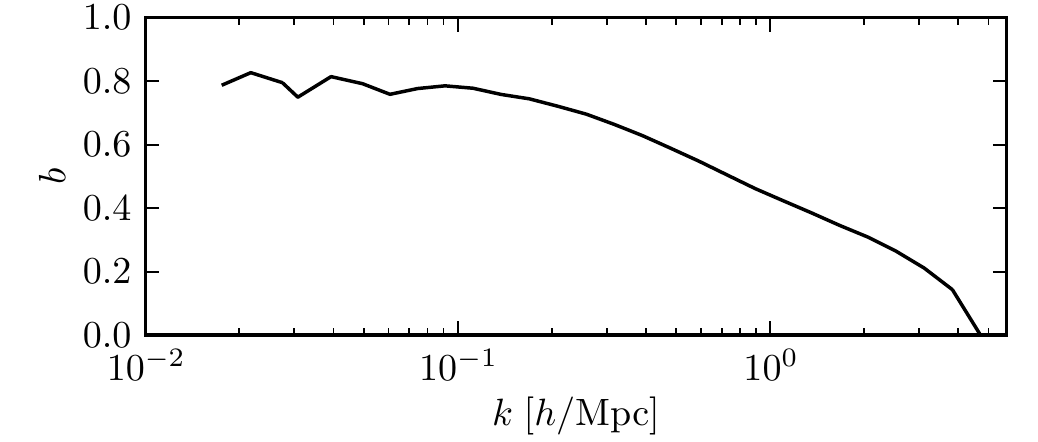}
        \caption{The halo bias parameter measured from S2 at $z = 0$.
          On scales $k \lesssim 0.2\ h/{\rm Mpc}$ the bias is roughly
          constant with $b \sim 0.8$. The bias falls off on smaller
          scales as power is suppressed within the typical virial
          radii of halos.}
        \label{fig:denbias}
      \end{center}
    \end{figure}
    
    The halo power spectrum is also plotted in Fig. \ref{fig:denpow}.
    As expected, the halo power follows the general shape of the dark
    matter power spectrum, but with a reduced amplitude, or bias.
    This bias is defined as:
    \begin{equation}
      b \equiv \sqrt{\frac{P_{hh}}{P_{cc}}},
      \label{eq:bias}
    \end{equation}
    and is plotted as a function of $k$ in Fig. \ref{fig:denbias}.
    The bias is roughly constant on large scales with $b \sim 0.8$ and
    falls off on small scales as the halo density field does not
    include contributions from the ``one-halo'' term describing the
    internal mass profile of halos
    \citep{bib:scherrer/etal:1991}. Hence, halo power is suppressed on
    scales comparable to the typical virial radii of halos which
    occurs at $k \sim 0.2\ h/{\rm Mpc}$ for the largest halos in the
    box.

  \end{subsection}

  \begin{subsection}{Velocity}
    
    Fig. \ref{fig:velslice} compares slices of dark matter, neutrino,
    and dark matter-neutrino relative velocity computed from the
    simulation particles as well as reconstructed from the dark matter
    and halo density fields using Eq. \ref{eq:vrec}. We observe a
    similar trend as the density fields with dark matter and neutrinos
    highly correlated in velocity. In addition, we see that the
    velocity fields reconstructed from only knowledge of either the
    dark matter or halo density field qualitatively agree with the
    large-scale structure of the velocity fields obtained within the
    simulation.

    \begin{figure*}[htbp]
      \begin{center}
        \includegraphics[width=0.9\textwidth]{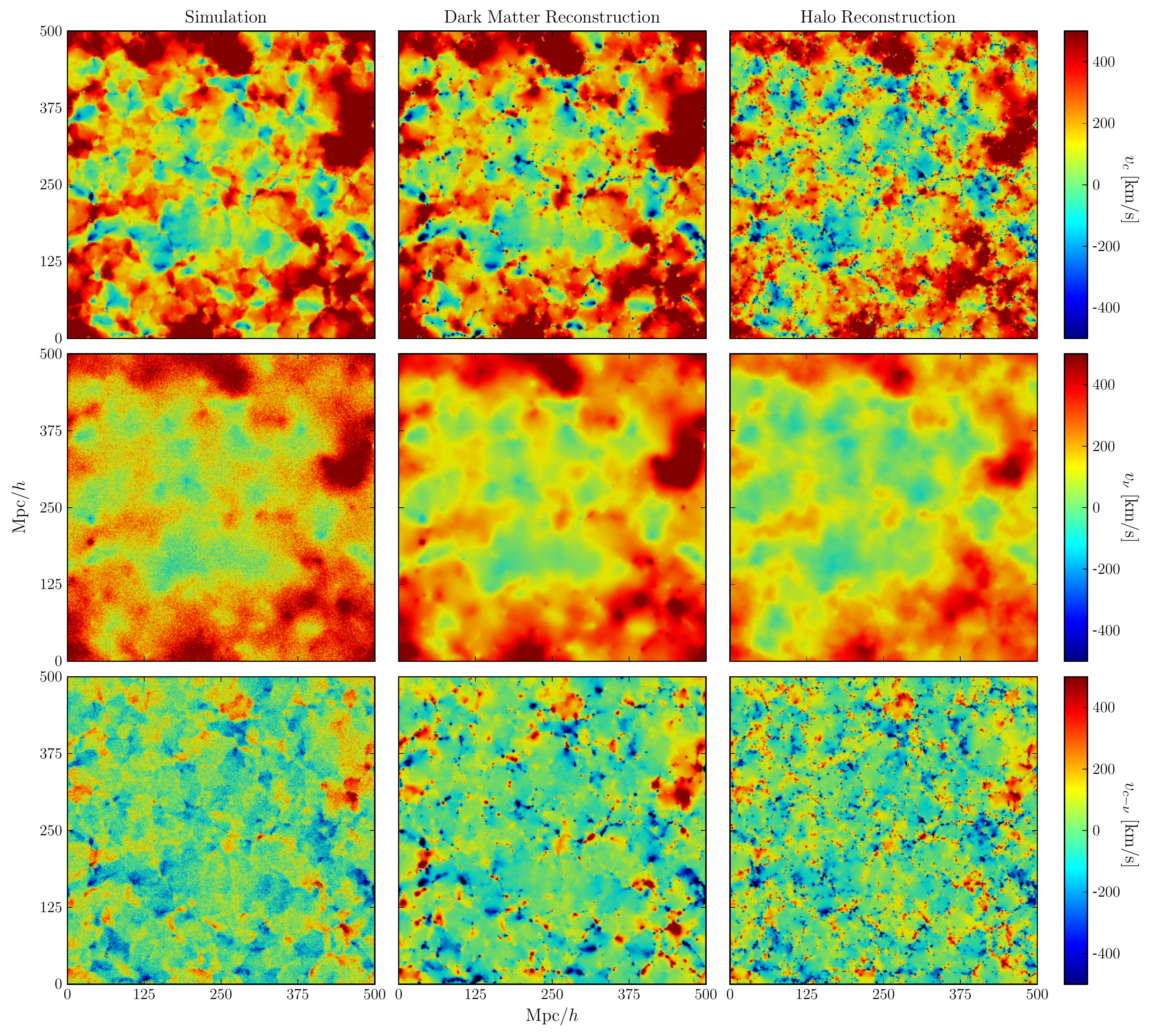}
        \caption{Slices of equal width $500\ \mpch{}$ and thickness
          $1.3\ \mpch{}$ showing the $z = 0$ velocity component
          perpendicular to the page for dark matter (top row),
          neutrinos (middle row), and the relative velocity between
          dark matter and neutrinos (bottom row). Columns show the
          velocity fields from the simulation particles (left column),
          reconstructed from the dark matter density field (middle
          column), and reconstructed from the halo density field
          (right column). We see that both of the reconstruction
          methods agree well with the large-scale structure of the
          simulation velocity fields.}
        \label{fig:velslice}
      \end{center}
    \end{figure*}

    Fig. \ref{fig:velpow} compares the simulated dark matter and
    neutrino velocity power spectra to the dark matter and halo
    reconstructed fields. Note that for the latter we take
    $\delta = \delta_h/b$ in Eq. \ref{eq:vrec} to account for the halo
    bias. We use a value of $b = 0.80$ consistent with the large-scale
    bias found in Fig. \ref{fig:denbias}.  We compute theoretical
    predictions for the velocity power using Eq. \ref{eq:PNL} with
    $T_i$ being a velocity transfer function.  We note that the groups
    method has also effectively removed shot noise from the velocity
    power just as for the density.
    
    Fig. \ref{fig:velpow} demonstrates that the simulated dark matter
    velocity field is suppressed on scales
    $0.2 \lesssim k \lesssim 4.0 \:\hmpc{}$ compared to the linear and
    non-linear expectations. This suppression was also seen in
    \citep{bib:Pueblas,bib:Hahn2014} and may be due to the
    thermalization of dark matter within collapsed objects.  The
    velocity field reconstructed from dark matter agrees well with the
    non-linear expectation of Eq. \ref{eq:PNL}. This is simply a
    reflection of the agreement between the dark matter density field
    and \hfit{} shown in Fig. \ref{fig:denpow}.  If we used the full
    bias curve, $b(k)$, instead of a constant then the halo
    reconstruction method works equally well.  Neutrinos, on the other
    hand, have a velocity power spectrum that agrees well with the
    non-linear expectation on scales $k\lesssim 0.15 \:\hmpc$.
    However, we find that they are underpredicted by linear theory on
    small scales.  It is unclear why neutrinos behave in an opposite
    manner from dark matter.

    \begin{figure}[htbp]
      \begin{center}
        \includegraphics[width=0.5\textwidth]{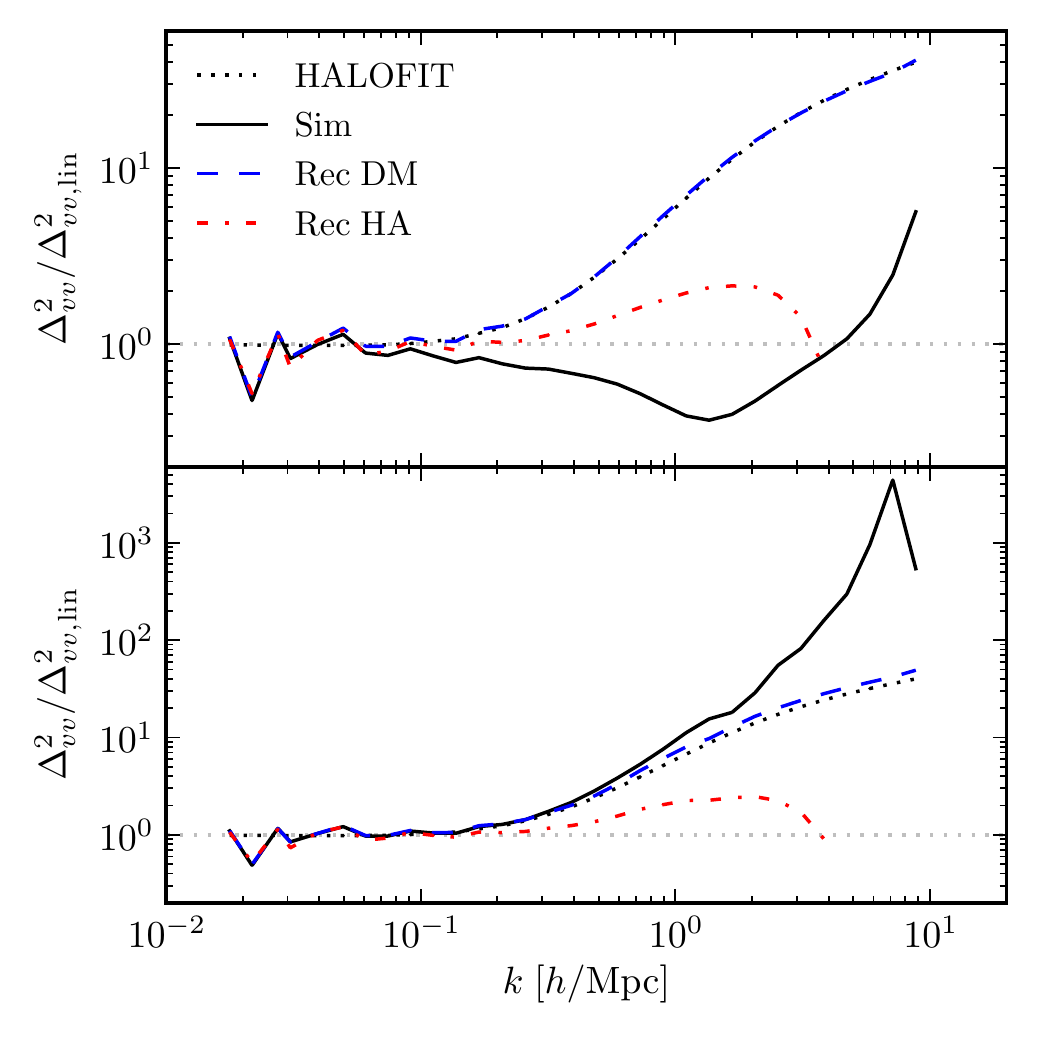}
        \caption{Velocity power spectra at $z = 0$ from S2 for dark
          matter (top) and neutrinos (bottom) normalized to the linear
          theory result obtained from Eq. \ref{eq:PNL}.  In each
          panel, the dotted black line shows the non-linear
          expectation of Eq. \ref{eq:PNL}, the solid black line shows
          the simulation result, and the dashed blue line (dot-dashed
          red line) shows the velocity field reconstructed from
          Eq. \ref{eq:vrec} using the dark matter (halo) density
          field.}
        \label{fig:velpow}
      \end{center}
    \end{figure}
   
    \begin{figure}[htbp]
      \begin{center}
        \includegraphics[width=0.5\textwidth]{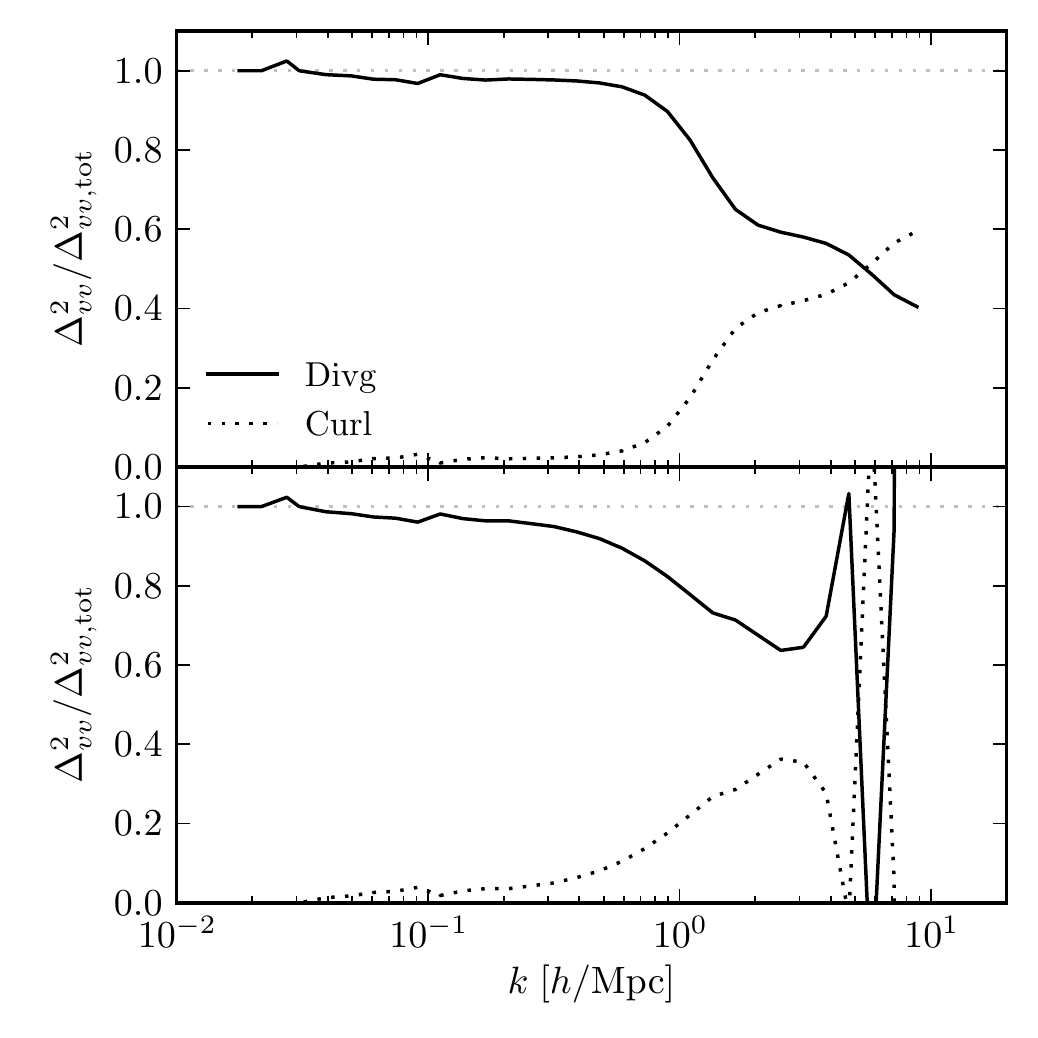}
        \caption{Relative fraction of the divergence (solid black
          line) and curl (dotted black line) components of the dark
          matter (top) and neutrino (bottom) velocity power at $z = 0$
          from S2. In each case, the curl component is negligible on
          scales $k \lesssim 1\ \hmpc{}$.  The oscillations seen with
          the neutrino power on small scales is indicative of their
          shot noise.}
        \label{fig:veldiv}
      \end{center}
    \end{figure}
   
    The efficacy of reconstructing velocities using Eq. \ref{eq:vrec}
    relies on the linearity of the velocity field. To test this we
    decompose velocity into divergence and curl components. We have
    performed this computation using both real-space finite
    differencing of the velocity field as well as Fourier space
    decomposition:
    \begin{eqnarray}
      \label{eqn:divcurl}
      \vec{v}_k &= \hat{k}(\hat{k}\cdot\vec{v}_k) + \hat{k}\times(\hat{k}\times\vec{v}_k) \nonumber \\
                &=\hat{k}D +\vec{C},
    \end{eqnarray}  
    where $D$ is the divergence field and
    $\vec{C} = \vec{v}_k - \hat{k}D$ is the curl field. Both the
    real-space and Fourier-space methods produce equivalent
    results. In linear theory, the velocity is parallel to $\hat{k}$
    and therefore has no curl. Hence, the presence of a curl component
    of the velocity field allows us to measure its degree of
    non-linearity.
 
    In Fig. \ref{fig:veldiv} we plot the divergence and curl
    components of both the dark matter and neutrino velocity
    fields. In each case, we see that the velocity is curl-free on
    scales $k \lesssim 1\ \hmpc{}$.  The only significant curl
    component occurs for dark matter on scales $k \gtrsim 5\ \hmpc{}$.
    This result highlights that the discrepancy between the simulated
    dark matter velocity and theoretical curves in
    Fig. \ref{fig:velpow} is not due to the presence of a curl
    component but rather due to non-linear processes affecting the
    divergence.

  \end{subsection}

  \begin{subsection}{Relative Velocity}
    \label{ssec:relvel}
    
    \begin{figure}[htbp]
      \begin{center}
        \includegraphics[width=0.5\textwidth]{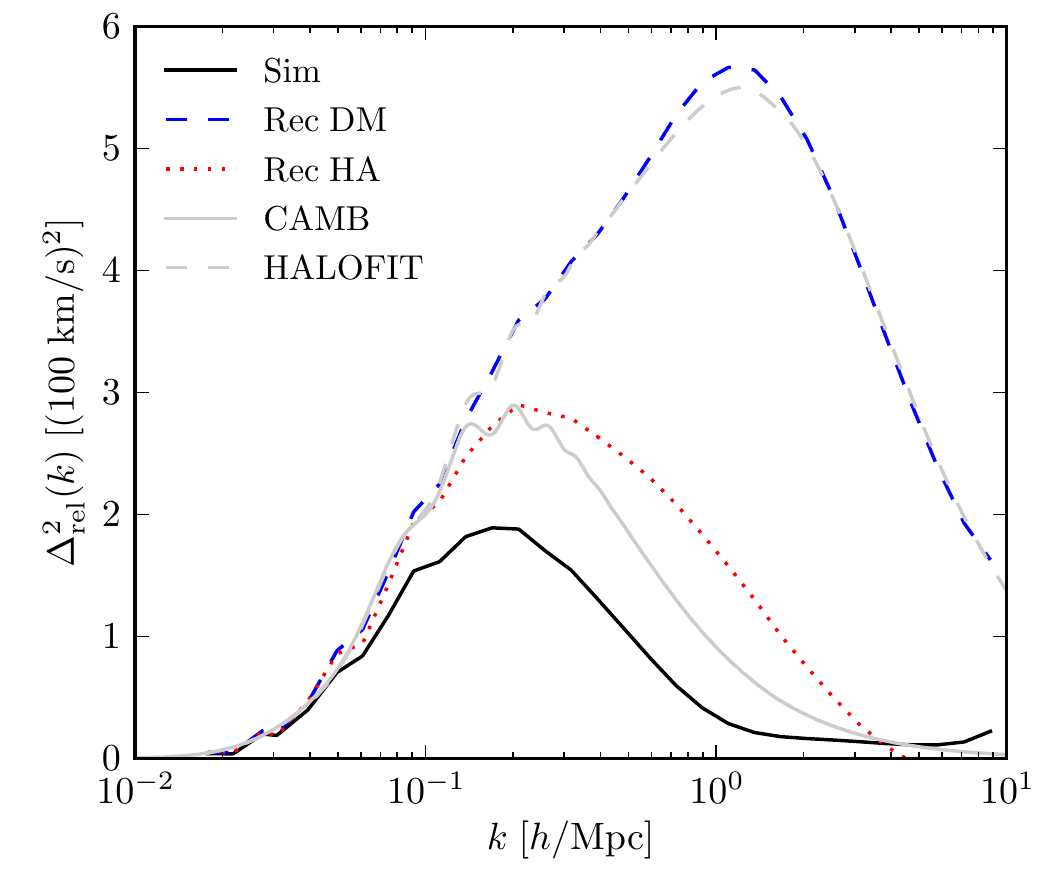}
        \caption{The dark matter-neutrino relative velocity power
          spectrum at $z = 0$ for S2 (black line) compared to the dark
          matter (dashed blue) and halo (dotted red) reconstructed
          fields as well as the linear (solid gray) and non-linear
          (dashed gray) predictions.  The simulated relative velocity
          power is similar to the linear prediction whereas the two
          reconstructed fields deviate from the linear curve due to
          non-linear structure formation.}
        \label{fig:relvelpow}
      \end{center}
    \end{figure}
    
    Fig. \ref{fig:relvelpow} compares the dark matter-neutrino
    relative velocity power spectrum to linear and non-linear
    predictions as well as to the two reconstruction methods. The
    relative velocity field from the simulations is roughly similar to
    the linear theory expectation, being within a factor of $3$ on
    scales $k<5\:\hmpc$.  The power spectra from the halo
    reconstruction method is also similar to the linear theory result.
    The field reconstructed from dark matter looks very different from
    the previous two but is consistent with the non-linear
    expectation. This can be made consistent with the linear theory
    result by simply multiplying Eq. \ref{eq:vrec} by the ratio
    between the linear and non-linear dark matter density power
    spectra.
   
    \begin{figure}[htbp]
      \begin{center}
        \includegraphics[width=0.5\textwidth]{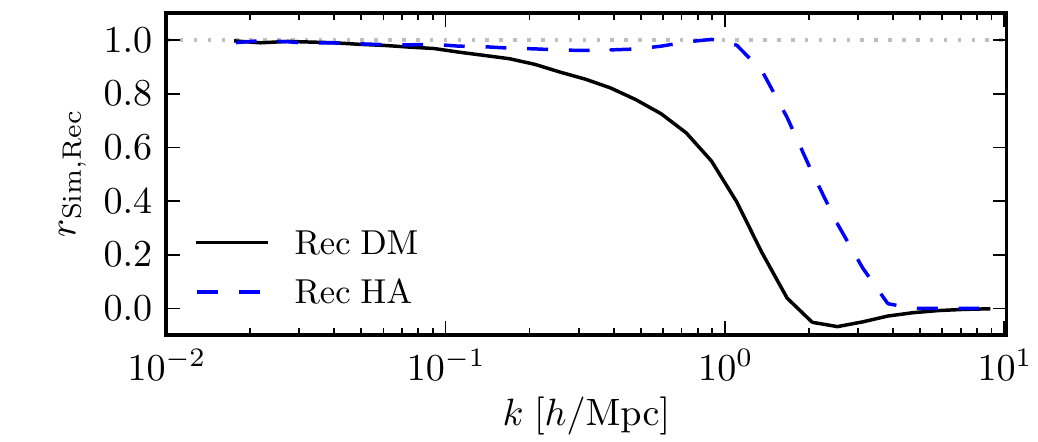}
        \caption{The dark matter-neutrino relative velocity
          correlation coefficient between the simulated field and the
          field reconstructed from dark matter (solid black line) and
          halo (dashed blue line) density fields. Both methods are
          highly correlated over all relevant scales.}
        \label{fig:relcorr}
      \end{center}
    \end{figure}
    
    Fig. \ref{fig:relcorr} shows the correlation coefficient defined
    in Eq. \ref{eq:rij} between the simulated and reconstructed
    relative velocity fields.  We see that both reconstruction methods
    reproduce the relative velocity field well over the scales of
    interest. In particular, the halo reconstruction achieves nearly
    perfect correlation on scales $k \lesssim 1\ \hmpc{}$.  The
    velocity correlation coefficient is a measure of how well the
    vector fields agree in direction as the denominator in
    Eq. \ref{eq:rij} divides out the magnitudes.  Thus, Fig.
    \ref{fig:relcorr} demonstrates that we are able to reconstruct the
    direction of the relative velocity field accurately.

    Fig. \ref{fig:allrelvelpow} shows the relative velocity power
    spectra for each of the four neutrino masses using the nearest
    particle/momentum method.  We find that they follow the same
    trends: lighter neutrinos have less relative velocity and the
    linear prediction is larger than in simulation.  Table
    \ref{tab:corrco} lists the integrated correlation coefficients as
    a function of neutrino mass between simulated and halo
    reconstructed velocities for dark matter, neutrino and dark
    matter-neutrino relative velocities.  We find that there is a
    large correlation between these methods indicating that the
    reconstruction method is accurately reproducing the simulation
    velocities.
    \begin{table}
      \caption{The integrated correlation coefficient defined in
        Eq. \ref{eq:intrij} between the simulated velocities and those
        reconstructed by halos for dark matter, neutrinos and dark
        matter-neutrino relative velocities.}
      \label{tab:corrco}
      \centering
      \begin{tabular} {c | c | c | c}
        \hline
        $m_\nu$ & Dark Matter & Neutrinos & Relative \\
        \hline
        0.05 & 0.95 & 0.98 & 0.94\\
        0.1 & 0.95 & 0.97 & 0.93\\
        0.2 & 0.95 & 0.97 & 0.91\\
        0.4 & 0.95 & 0.97 & 0.88\\
        \hline
      \end{tabular}
    \end{table}
  
    \begin{figure}[htbp]
      \begin{center}
        \includegraphics[width=0.5\textwidth]{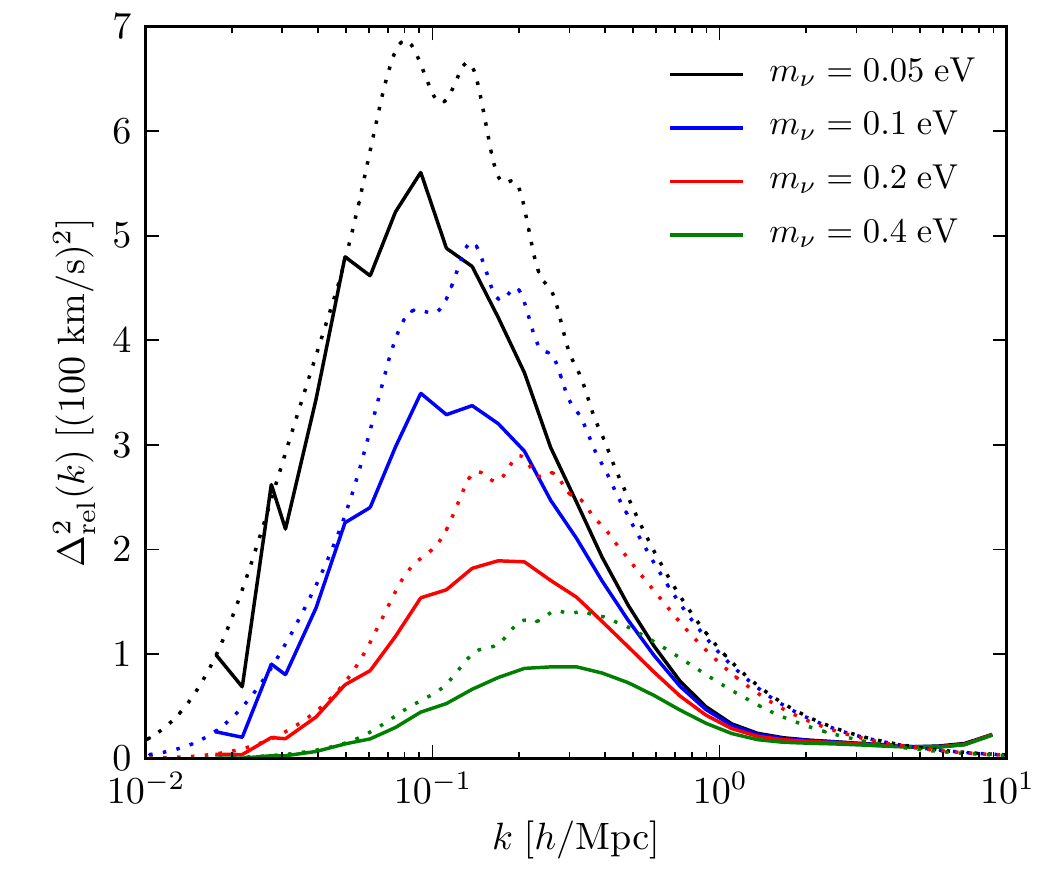}
        \caption{The dark matter-neutrino relative velocity power
          spectra via the nearest particle/momentum method for all
          four neutrino masses (solid) along with theoretical
          predictions (dashed).  The power is clearly suppressed
          compared to linear theory but behaves qualitatively similar
          with varying masses.}
        \label{fig:allrelvelpow}
      \end{center}
    \end{figure}

    Finally, in Fig. \ref{fig:corrlength} we show the relative
    velocity correlation lengths, $\xi_{1/2}$, defined as in
    \citep{bib:Zhu2013} to be the point at which the relative velocity
    correlation function,
    \begin{equation}
      \xi_{\nu c}(r) = \int \frac{dk}{k} \Delta^2_{\nu c}\frac{\sin(kr)}{kr}
      \label{eq:corrfun}
    \end{equation}
    reaches half its maximum value. This scale can be thought of as
    the size of a region with a uniform velocity field.  Lighter
    neutrinos are less affected by large scale structure due to their
    larger thermal velocities and so are coherent over larger regions.
    Fig. \ref{fig:corrlength} shows these correlation lengths as a
    function of neutrino mass.  We find that the simulations exhibit a
    slightly larger correlation length for each neutrino mass compared
    to the theoretical predictions.  The shapes of the curves remain
    similar, however, with both having power law slope which we fit to
    have an exponent $-0.44$.

    \begin{figure}[htbp]
      \begin{center}
        \includegraphics[width=0.5\textwidth]{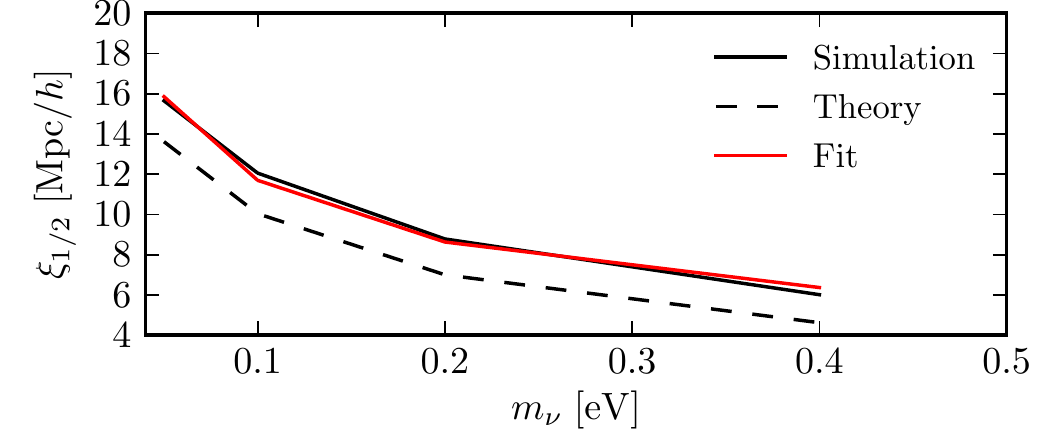}
        \caption{The correlation length defined to be the distance for
          which the correlation function in Eq. \ref{eq:corrfun} drops
          to half its maximum value for varying neutrino masses.  The
          simulations have longer correlation lengths but follow a
          similar power law behaviour.}
        \label{fig:corrlength}
      \end{center}
    \end{figure}

  \end{subsection}

\end{section}

\begin{section}{Discussion}
  \label{sec:discussion}

  We have tested four methods of computing the velocity field: a
  nearest particle method, a momentum method, and reconstruction via
  dark matter and halo density fields.  Our results are generally
  consistent with theoretical expectations and highly correlated among
  each other.  Specifically we have demonstrated that reconstructing
  the velocity from point-particle halos produces a velocity field
  highly correlated with that of our N-body particles.  It is the near
  unit correlation coefficient - a measure of the angle between the
  two fields - that ensures that the reconstructed velocity points in
  the right direction.  The magnitude of the velocity can then simply
  be scaled to the correct value as long as the bias is known.

  This result allows for a prescription to determine the actual
  velocity fields in our own Universe.
  \begin{enumerate}
  \item Reconstruct the galaxy density field from a galaxy survey
    catalogue.  We expect this reconstruction to be very comparable to
    the halo reconstruction we use here except with the addition of a
    1-halo term to make the bias constant over more wavenumbers.
  \item Fourier transform the gridded density field.  Then, use
    Eq. \ref{eq:vrec} to compute the dark matter, neutrino and
    relative velocity fields in Fourier space.  Here, a non-linear
    correction can be applied by additionally multiplying by a factor
    of $\Delta_v^{\rm Sim}/\Delta_v^{\rm Rec HA}$.
  \item Fourier transform back to real space.
  \end{enumerate}
  We first note that a similar process could be performed on the
  density fields produced by 21 cm observations.  We also note that
  redshift distortions and masking effects might result in extra
  biases in the reconstruction scheme.  We intend to investigate these
  effects in a future paper.

  Our results also provide support for the applicability of the
  analysis performed in \citep{bib:Zhu2013,bib:Zhu2014}.  They used
  moving background perturbation theory to study the neutrino relative
  velocity effect.  The moving background approximation relies on
  having a coherent background relative flow and our simulation
  results indicate that the coherency scales of such motions are
  larger than predicted.  Thus, we expect that inaccuracies in the
  predicted dipole distortion to the correlation function will come
  from non-linear evolution rather than the moving background
  approximation.  We note that we can also directly measure the dipole
  correlation function in our simulations and plan to report on this
  in a subsequent paper.

\end{section}

\begin{section}{Conclusion}
  \label{sec:conclusion}

  We performed a set of four large N-body simulations including cold
  dark matter and neutrinos of varying mass.  We have accurately
  measured the dark matter-neutrino relative velocity.  We find that
  we can accurately reconstruct this velocity using a linear theory
  approach and halo density fields.  We have described a simple method
  for accurately predicting the relative velocity field via a galaxy
  survey or 21 cm observations.  Since such a reconstruction allows
  for an independent measurement of neutrino masses, we expect this
  technique to provide significant constraints in upcoming
  astronomical surveys.

\end{section}

\begin{section}{Acknowledgements}
  \label{sec:acknowledgements}

  We acknowledge valuable discussions with Joel Meyers and Yu Yu.  We
  acknowledge the support of NSERC.
  
  Computations were performed on the GPC supercomputer at the SciNet
  HPC Consortium \citep{bib:Loken}. SciNet is funded by: the Canada
  Foundation for Innovation under the auspices of Compute Canada; the
  Government of Ontario; Ontario Research Fund - Research Excellence;
  and the University of Toronto.

\end{section}

\end{document}